\documentclass[conference]{IEEEtran}
\IEEEoverridecommandlockouts

\usepackage{graphicx}
\usepackage{subcaption}
\usepackage{placeins}
\usepackage{amsmath}
\usepackage{cite}
\usepackage{url}

\begin{document}

\title{Frequency Lock Encoding}

\author{\IEEEauthorblockN{Joshua Montierth, Michael Rice, Philip Lundrigan}
    \IEEEauthorblockA{\textit{Department of Electrical and Computer Engineering} \\
        \textit{Brigham Young University}\\
        Provo, Utah, USA \\
        \{jmontierth, mdr, lundrigan\}@byu.edu}
}

\maketitle

\begin{abstract}
    Modern wireless systems are designed with excess synchronization bandwidth to ensure reliable operation under worst-case conditions. This paper presents a protocol-agnostic secondary signaling layer that exploits this unused synchronization margin by intentionally introducing small, controlled frequency offsets into an existing communication signal to embed a parallel stream of information. These offsets are naturally absorbed by the primary receiver’s carrier frequency offset (CFO) correction mechanisms, allowing the primary data stream to remain intact while enabling simultaneous secondary communication. The proposed signaling method employs a frequency-shift-keyed overlay engineered to remain within the stable operating region of conventional carrier-synchronization loops, ensuring backward compatibility and transparency for legacy receivers. We analyze the system for single-carrier protocols, characterize the trade-offs between secondary throughput and the impact on the primary link, and validate performance through both simulation and over-the-air experiments using software-defined radios. Results demonstrate reliable secondary communication with negligible degradation to primary performance, making our system well-suited for ad hoc spectrum sharing and decentralized coordination at the tactical edge.
\end{abstract}

\begin{IEEEkeywords}
    tactical communications, spectrum sharing and coordination, subprotocol, phase-locked loop, software-defined radio
\end{IEEEkeywords}
\section{Introduction}

Wireless links face unpredictable analog disturbances that introduce errors into transmitted signals. To ensure reliability under worst-case conditions, protocols are designed with powerful synchronization mechanisms that reserve additional bandwidth to tolerate impairments. Because failure is often catastrophic, designers consistently provision systems for worst-case operating conditions. As a result, modern wireless systems contain a hidden inefficiency: \textit{excess impairment correction capacity} under many operating environments. While this safety margin guarantees robustness, it also means that valuable bandwidth is frequently spent protecting against errors that are not actually present.

In this paper, we propose a different perspective on signal impairments and noise in wireless communication. Instead of treating unused synchronization capacity as wasted overhead, we intentionally and opportunistically introduce \textit{controlled, meaningful impairments} into a signal to create a secondary layer of communication without disrupting the primary communication. To the primary protocol, these perturbations appear as ordinary noise; however, to a receiver that knows how to interpret them, they represent a \textit{second parallel channel of information}. This secondary channel can support a fully featured ``subprotocol'' capable of carrying coordination data, such as in-band signaling between unassociated networks without prior coordination or spectrum usage information for agile spectrum operations. In this way, what would otherwise be unused correction capacity becomes a new signaling resource that enables simultaneous communication \textit{without requiring additional spectrum, antennas, or modifications to the underlying protocol}.

This approach is powerful for three reasons. First, it is protocol-agnostic. Because the subprotocol operates by embedding information within controlled perturbations that are corrected by the primary receiver's synchronization algorithms, it does not depend on the modulation, framing, or higher-layer structure of the underlying protocol. Second, it is backward compatible and, as a result, can potentially be used with existing commercial technologies. A legacy receiver that is unaware of the subprotocol simply interprets the intentional perturbations as ordinary noise. Third, it is opportunistic. The subprotocol consumes only the excess impairment-correction capacity available in the primary link. When the full synchronization capability of the primary system is required, the secondary signaling can be disabled entirely without affecting normal operation. These properties make the proposed approach well-suited for decentralized and agile spectrum coordination and resilient edge communications.

\begin{figure}
    \centering
    \includegraphics[width=0.90\linewidth]{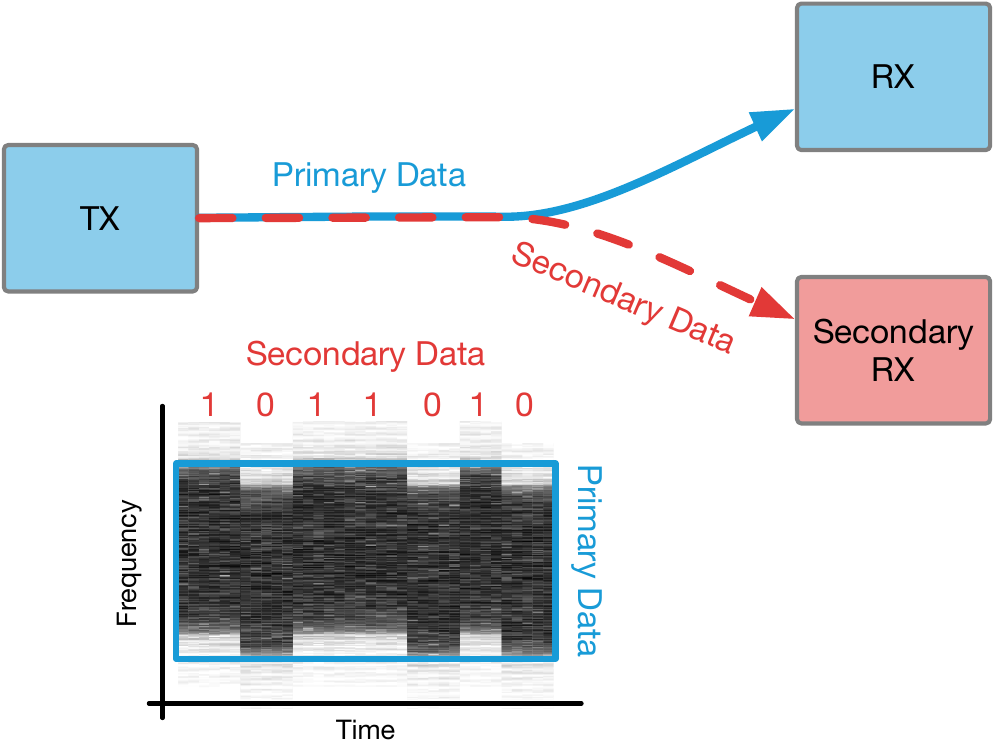}
    \caption{By intentionally introducing a controlled frequency offset into the primary signal, secondary spectrum coordination information can be conveyed simultaneously.}
    \label{fig:overview}
\end{figure}

Although many forms of meaningful impairments could be used to realize such a subprotocol, this work focuses specifically on embedding information through controlled frequency offsets applied to the underlying communication waveform, as illustrated in Fig.~\ref{fig:overview}. The choice of frequency perturbations over other synchronization domains, such as timing, is deliberate. Frequency correction feedback loops generally do not suffer from self-noise in the same manner as timing recovery loops, resulting in a cleaner observable signal. Additionally, while frequency synchronization is often implemented in hardware, the outputs of these algorithms are commonly exposed to firmware or software layers, unlike many synchronization mechanisms in other domains.

Others have used similar techniques to provide covert communication~\cite{secret_agent_radio, 10360170, 10.1145/3531536.3532959, classen_practical_2015}, which we will discuss in the next section. However, the primary goal of this paper is to establish the mathematical foundation for such a subprotocol and to demonstrate its practical feasibility for spectrum coordination through preliminary simulations and software-defined radio (SDR) experiments. To this end, we make the following contributions:

\begin{itemize}
    \item We introduce the concept of encoding a secondary stream of information into a primary transmission through controlled variation of the carrier frequency offset, which we call \textbf{Frequency Lock Encoding} (FLE). Although the method is protocol-agnostic, we develop and evaluate it in the context of single-carrier waveforms.

    \item We develop the mathematical framework necessary to describe and analyze FLE, and characterize its behavior across a range of operating parameters. We show that, under appropriate conditions, FLE introduces little to no measurable degradation to the primary modulation scheme.

    \item We implement a proof-of-concept realization using software-defined radios (SDRs) and perform preliminary over-the-air experiments demonstrating that the approach is practically achievable while achieving data rates of over 900~kbps.
\end{itemize}
\section{Related Work}

\subsection{Covert Communication}

The idea of leveraging waveform parameters to embed covert data has been well studied for many years~\cite{secret_agent_radio, 10360170, 10.1145/3531536.3532959, classen_practical_2015}. Such approaches range from embedding data in constellations by emulating noisy channel conditions~\cite{secret_agent_radio} to emulating hardware noise using machine learning~\cite{10360170}. The previous work most similar to our work was done by Classen et al.~\cite{classen_practical_2015}, where the authors explore multiple covert channels for Wi-Fi systems, one of which uses carrier frequency offsets. However, their approach focuses only on Wi-Fi, the frequency offsets are much larger (1--50~kHz), and, as a result, negatively affect the primary Wi-Fi communication. Our approach is not tied to Wi-Fi and primarily focuses on single-carrier waveforms. The frequency offsets used in FLE are much smaller, making our approach more spectrum efficient and less detrimental to the primary protocol. Our work aims to provide a mathematical framework for how to think about using frequency offsets to embed data. While FLE could be used for covert communication, we focus on coordination rather than covertness, and as a result, the goals of our system significantly differ: we are more concerned with not disturbing the primary communication. We have also developed a fully featured and usable subprotocol, with its own framing and error detection, whereas others have focused on avoiding detection.

\subsection{Spectrum Coordination}

Large-scale spectrum coordination is often performed using a centralized database, such as those used by CBRS~\cite{noauthor_spectrum_nodate} and Wi-Fi 6E~\cite{afc}. However, single points of failure and dependence on backhaul connectivity are poorly suited to tactical operations in denied or disconnected environments. Our approach enables peer-to-peer spectrum coordination without requiring a centralized database.

Other ad hoc spectrum-sharing protocols have been explored. Similar to our approach, they modify transmission characteristics to embed spectrum-sharing data. These methods include embedding spectrum data in an OFDM subcarrier~\cite{weldegebriel_sensing_2025}, in the amplitude of a signal~\cite{9795976}, in a sequence of transmissions using on-off keying~\cite{palacios_network_2024}, and in inter-transmission timing~\cite{palacios_hidden_2025}. While similar in goal, our approach differs substantially from these works. We use frequency offsets to encode secondary data, yielding a much higher data rate (hundreds of kbps compared to tens of bps).

\section{System Design}

In developing FLE, we had the following two constraints: avoid degrading the primary communication while sending as much data as possible through the secondary channel. To this end, we explore the mathematical bounds of such a system in the context of a single-carrier system and describe how it can be used for spectrum coordination. Much of the analysis could be applied to multi-carrier systems as well, but that is beyond the scope of this paper.

\subsection{Overview}

\begin{figure*}
    \centering
    \includegraphics[scale=0.52]{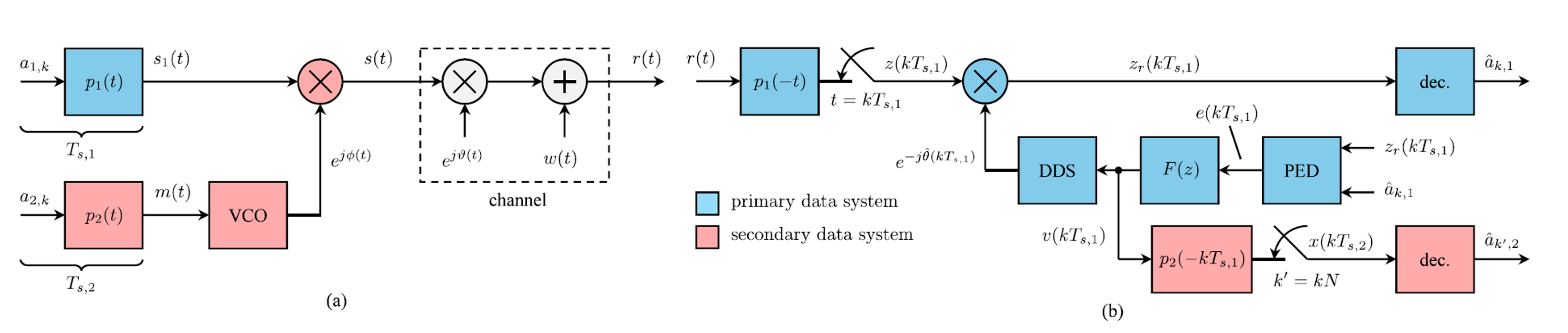}
    \caption{A block diagram illustrating the FLE technique for single-carrier modulation:
        (a) the transmitted signal $s(t)$ and the received signal $r(t)$;
        (b) the FLE detector (the loop filter output of the primary data serves as the input for secondary data detection).}
    \label{fig:mdr-system2}
\end{figure*}

The signals generated by the primary and secondary systems are illustrated in Fig.~\ref{fig:mdr-system2}.
The complex-valued lowpass equivalent \cite{proakis:2008} of the primary signal is
\begin{equation}
    s_1(t) = \sum_k a_{1,k} p_1(t-kT_{s,1})
    \label{eq:primary1}
\end{equation}
where $a_{1,k}$ is the $k$-th QAM symbol (or constellation point), $p_1(t)$ is the pulse shape, and $T_{s,1}$ is the symbol time.
The secondary signal begins with a pulse amplitude modulation (PAM) pulse train
\begin{equation}
    m(t) = \sum_{k'} a_{2,k'} p_2(t-k'T_{s,2})
    \label{eq:secondary1}
\end{equation}
where $a_{2,k'}$ is the $k'$-th real-valued PAM symbol, $p_2(t)$ is the pulse shape, and $T_{s,2}$ is the symbol time.
The PAM pulse train $m(t)$ is frequency modulated to produce the complex-valued lowpass equivalent signal $s_2(t) = e^{j\phi(t)}$ where
\begin{align}
    \phi(t) & = 2\pi\Delta f \int_{0}^{t} m(u) du
    \\
            & = 2\pi\Delta f\sum_{k'} a_{2,k'} q_2(t-k'T_{s,2})
    \\
    q_2(t)  & = \int_{0}^{t} p_2(u) du
\end{align}
The complex-valued lowpass equivalent of the transmitted signal using FLE is
\begin{equation}
    s(t) = s_1(t) e^{j\phi(t)}
\end{equation}
The phase variations in $s(t)$ are due to the secondary transmission.
The modulation index $\Delta f$ should be small enough to allow the PLL at the detector to track out the phase variations.

The received signal is
\begin{equation}
    r(t) = s_1(t) e^{j\theta(t)} + w(t)
    \label{eq:r1}
\end{equation}
where
\begin{equation}
    \theta(t) = \vartheta(t) + \phi(t)
    \label{eq:r2}
\end{equation}
where $\vartheta(t)$ is the phase shift due to the propagation medium
and  $w(t)$ is the thermal noise modeled as a white wide-sense stationary complex-valued normal random process with zero mean and power spectral density $2N_0$ W/Hz \cite{proakis:2008}.
A block diagram of the channel is illustrated in Figure~\ref{fig:mdr-system2}(a).

The detectors for the primary and secondary data are shown in Figure~\ref{fig:mdr-system2}(b).
The received signal is applied to a filter matched to the primary pulse shape $p_1(t)$ and sampled every $T_{s,1}$ seconds in synchronism with the symbol transitions on the primary data channel.
(A symbol timing synchronizer is assumed, but not shown.)
The matched filter outputs form a discrete-time sequence that is applied to a decision-directed carrier phase PLL.
The $k$-th matched filter output $z(kT_{s,1})$ is derotated by the $k$-th phase estimate $\hat{\theta}(kT_{s,1})$  to produce $z_r(kT_{s,1})$.
The phase error detector (PED) produces the $k$-th error signal $e(kT_{s,1})$ based on the maximum likelihood principle (see \cite{rice:2018}).
The error signal $e(kT_{s,1})$ is applied to a loop filter with $z$-domain transfer function $F(z)$ to produce $v(kT_{s,1})$.
The loop filter output is applied to a direct digital synthesizer (DDS) to produce the phase estimate used for de-rotating the next matched filter output.

When in lock, the loop filter output $v(kT_s)$ is a noisy version of $T_{s,2}$-spaced samples of the derivative of $\theta(t)$:
\begin{multline}
    v(kT_{s,1}) \approx \vartheta'(kT_{s,1})
    + 2\pi \Delta f \sum_{k'} a_{2,k'} p_2(kT_{s,1}-k'T_{s,2}) \\
    + \text{noise}
    \label{eq:r3}
\end{multline}
The relationship \eqref{eq:r3} applies only when the bandwidth of the secondary pulse train is less than the closed-loop bandwidth of the PLL.
Assuming this condition, the loop filter output is a noisy version of the secondary pulse train plus the derivative of the phase shift induced by the propagation medium.
The following three scenarios present no problem to detecting the secondary signal:
\begin{enumerate}
    \item If the propagation medium induces a static phase shift, then $\vartheta(t) = \vartheta$ whose time derivative is zero.
    \item If the propagation medium induces a static frequency offset, then $\vartheta(t)$ is the phase ramp $\vartheta(t) = \Delta\omega t$ whose time derivative is the constant $\Delta\omega$.
          This constant may be removed from the sample sequence $v(kT_{s,1})$ using a DC blocker.
    \item If the propagation medium induces a slowly varying phase shift that can be considered quasi-static, then the derivative of $\vartheta(t)$ is small relative to the secondary pulse train.
\end{enumerate}
Assuming any one of these conditions, the loop filter output comprises $T_{s,1}$-spaced samples of a noisy version of the secondary PAM pulse train.
Thus, detection may be accomplished by applying $v(kT_{s,1})$ to a discrete-time FIR filter matched to $T_{s,1}$-spaced samples of the secondary pulse shape.
Every $N = T_{s,2}/T_{s,1}$-th sample of the matched filter output is used for detection of the secondary symbols.
The optimum phasing of the downsampling operation is controlled by a symbol-timing synchronizer (not shown).

\subsection{Design Constraints and Parameters}

The description in the previous section identified the key design constraints:
\begin{itemize}
    \item The modulation index $\Delta f$ or its normalized version $\Delta f T_{s,1}$.
          The modulation index controls the magnitude of the frequency deviation induced by the secondary PAM pulse train.
          On the one hand, if the $\Delta f$ is too large, then the primary signal's PLL cannot track it out; the performance of the primary signal is compromised.
          On the other hand, because the signal-to-noise ratio at the loop filter output is proportional to the product $(\Delta f T_{s,1})^2$ and the value of $E_b/N_0$ for the primary signal, reducing $\Delta f T_{s,1}$ too much compromises the performance of the secondary signal.
    \item The secondary signal data rate $1/T_{s,2}$.
          What is important is the secondary signal data rate relative to the primary signal data rate.
          This is represented by the ratio
          \begin{equation}
              N = \frac{T_{s,2}}{T_{s,1}} \quad\text{primary symbols/secondary symbols}.
          \end{equation}
          The throughput of the secondary channel increases with decreasing $N$.
          However, if $N$ is too small, then the bandwidth of the secondary pulse train is larger than the closed-loop bandwidth of the carrier phase PLL used by the primary system.
          This compromises the ability of the PLL to track out the phase variations---a negative impact on the performance of the primary system---and violates the assumptions on which the relationship \eqref{eq:r3}---a negative impact on the performance of the secondary system.
\end{itemize}

\subsection{System Feasibility}

While direct access to raw IQ samples or internal PLL signals is typically unavailable in commercial off-the-shelf (COTS) hardware (though not unheard of~\cite{ath10k}), this work explores the capabilities that could be realized if such signals were more readily exposed at the hardware level. These signals already exist within the digital logic of modern radios; in many cases, only register access would need to be propagated through the software stack. Similarly, applying an FSK overlay via carrier mixing is generally infeasible in software alone, but modest hardware extensions could enable this functionality on most COTS designs. A key insight is that although minor architectural changes would be required, they would remain \textbf{fully backward compatible with existing communication protocols}.



\subsection{Spectrum Coordination}

This system is particularly well-suited for spectrum coordination where radios from different units may operate in close proximity without shared network association or reliable access to centralized spectrum-management infrastructure. Because FLE embeds coordination information directly into the carrier-frequency dynamics of an ongoing transmission, a receiver \textit{need not demodulate or decode the primary payload to extract coordination information}. Instead, it must only synchronize to the signal sufficiently to estimate the CFO. This property is valuable in tactical settings where payload formats, encryption, or network credentials may differ across systems, but where basic physical-layer observability remains possible. 
We envision FLE being used to exchange low-rate coordination information such as frequency plans, channel-occupancy summaries, and network configuration information, thereby enabling lightweight peer-to-peer spectrum awareness without requiring prior coordination, protocol interoperability, or dependence on vulnerable backhaul links. As with any coordination channel, adversarial deployments require security measures such as message authentication and replay protection. These measures can be implemented above the FLE physical layer and are beyond the scope of this work.




\begin{figure*}[t]
    \centering
    \setlength{\tabcolsep}{2pt}
    \renewcommand{\arraystretch}{1.0}
    \begin{tabular}{cc}
        \includegraphics[width=0.43\textwidth]{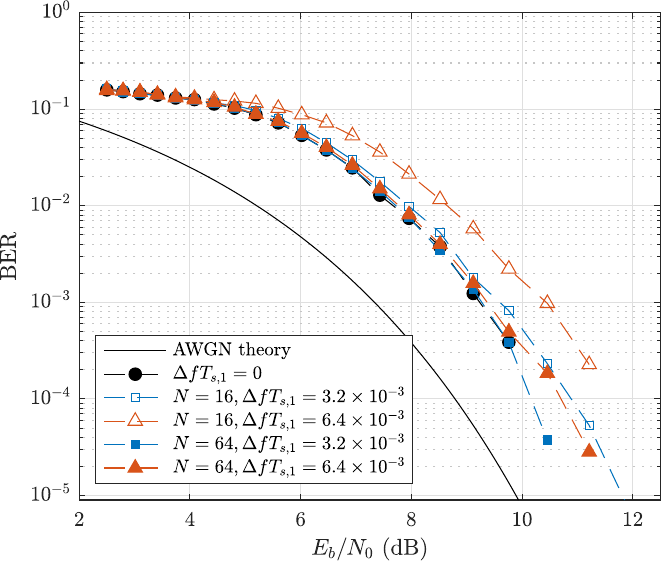}
        &
        \includegraphics[width=0.43\textwidth]{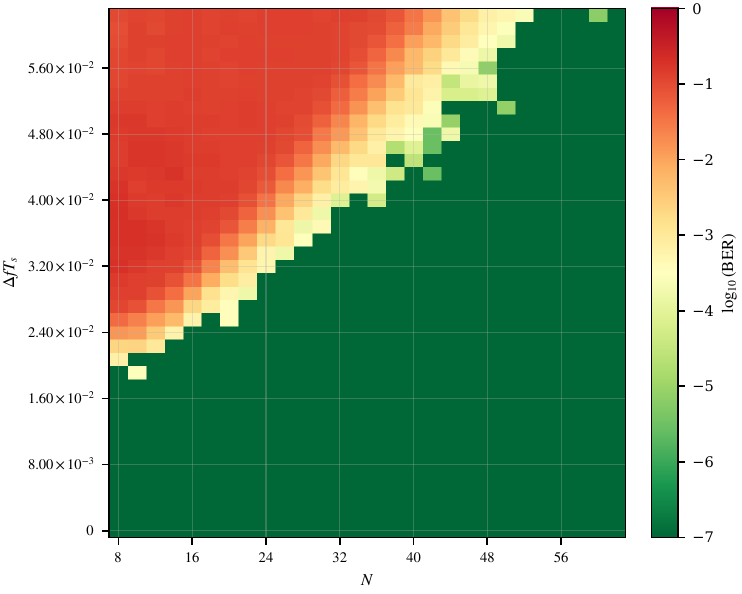}
        \\
        (a) & (b)
        \\
        \includegraphics[width=0.43\textwidth]{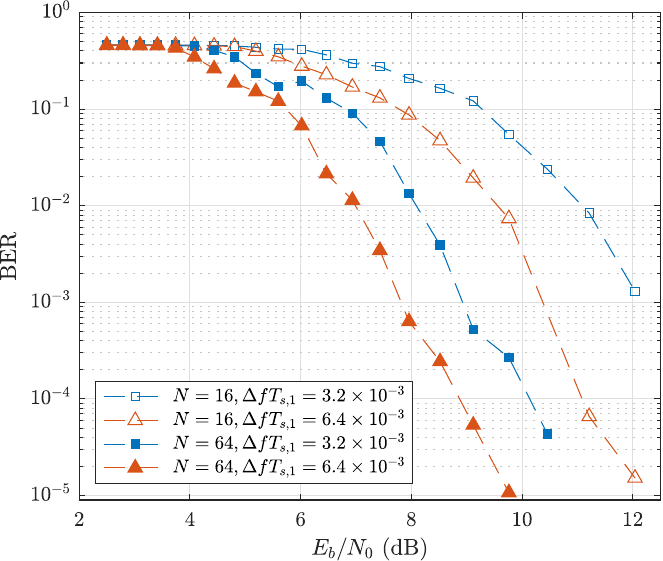}
        &
        \includegraphics[width=0.43\textwidth]{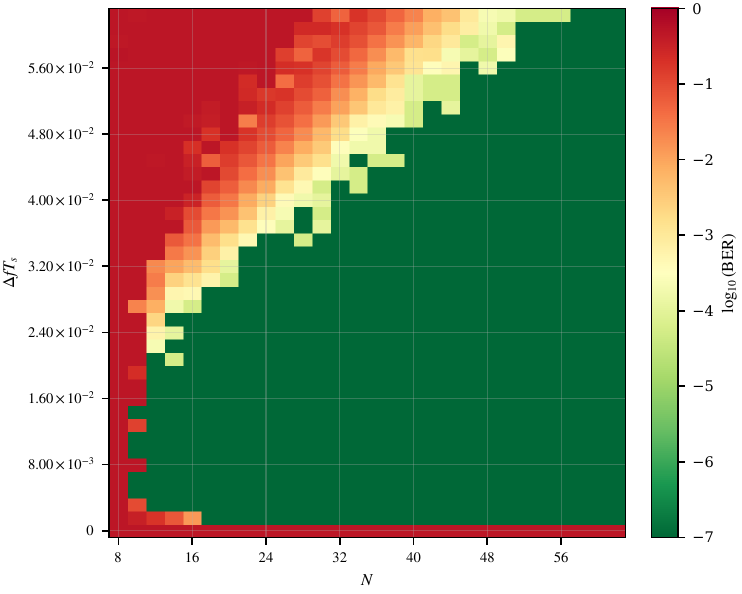}
        \\
        (c) & (d)
    \end{tabular}
    \caption{Single-carrier QPSK performance under Frequency Lock Encoding (FLE):
    (a) primary channel BER for representative FLE parameter settings;
    (b) primary channel BER in the $N$--$\Delta f T_s$ parameter space for $E_b/N_0 = 18$~dB;
    (c) secondary channel BER for representative FLE parameter settings;
    (d) secondary channel BER in the $N$--$\Delta f T_s$ parameter space for $E_b/N_0 = 18$~dB.}
    \label{fig:qpsk-results}
\end{figure*}

\section{Implementation of Real-Time System}

To demonstrate the performance of the single-carrier FLE, we implement our real-world pipeline in GNU Radio Companion. The primary signal comprised QPSK with the square-root raised-cosine pulse shape (SRRC) with unity DC gain and 50\% excess bandwidth, and the secondary signal was a binary PAM pulse train using the SRRC pulse shape with 100\% excess bandwidth.
For the computer simulation, AWGN noise was added using GNU Radio's channel model block.
For the primary channel, symbol timing synchronization was performed using the Gardner timing error detector \cite{rice:2018} and a Farrow interpolator for timing adjustments.
Carrier phase synchronization was performed using a second-order PLL with a closed loop equivalent bandwidth $B_nT_{s,1} = 0.02$ cycles/symbol.
Secondary channel detection was performed using the matched filter followed by a second symbol timing synchronizer using the Gardner timing error detector and Farrow interpolator.
Experiments were conducted using a variety of values for $\Delta f T_{s,1}$ and $N$, as summarized in Section~\ref{sec:evaluation}.




With the same GNU Radio Companion environment used in simulation, we transmit our FLE overlay over the air and decode it in real time. In this setup, both the primary and FLE signals used a packetized structure with a 32-bit sync word marking the beginning of the packet and a 32-bit CRC to ensure signal integrity. 
We use two USRP B210 SDRs as the transmitter and receiver of the primary protocol. 
The transmitter is configured to concurrently transmit FLE data. We use the ADALM-Pluto SDR as a secondary RX, which passively monitors the transmission and extracts only the FLE information embedded in the frequency offsets. Using the relatively inexpensive Pluto SDR shows the low barrier to entry for our protocol and the universality of this design across different hardware capabilities. For our over-the-air experiments, we transmit in the unlicensed 915~MHz band. This real-time over-the-air setup validates the real-world capabilities of FLE to enable spectrum sharing and coordination.

\section{Evaluation}
\label{sec:evaluation}

In our evaluation, we seek to answer three main questions: (1)~How much does FLE affect the performance of the primary communication link? (2)~Under what conditions can the secondary channel be reliably decoded? (3)~How do the FLE parameters $\Delta f T_s$ and $N$ affect system performance? Performance is measured using bit-error rate (BER) as a function of $E_b/N_0$ for both the primary and secondary protocols, calculated using the bit energy of the primary protocol.

\subsection{Simulation}

After extensive simulation, we find that although FLE parameters can influence QPSK performance, their impact on the primary protocol remains minimal provided the parameters lie within a defined operating envelope. Fig.~\ref{fig:qpsk-results}(a) shows the BER performance of QPSK under four different FLE configurations, along with a baseline case in which FLE is absent. For reference, the theoretical BER curve under additive white Gaussian noise (AWGN) is also included. The observed implementation loss is attributable to the simulation tooling rather than the presence of FLE. Nearly all simulated curves, including the no-FLE case, follow the same trajectory. Only when parameters are pushed to extreme values (e.g., $N=16$ and $\Delta f T_s=0.0064$) does a slight degradation appear. Even in this case, the degradation is limited to approximately 1~dB in $E_b/N_0$.

To identify the bounds of this stability envelope, we perform a parameter sweep, shown in Fig.~\ref{fig:qpsk-results}(b). Primary-protocol reception remains near-perfect below a clear boundary in the $N$--$\Delta f T_s$ plane, indicating a wide operating region where FLE can coexist with QPSK without significant impact.

In contrast, the effect of FLE parameters on the subprotocol is more prominent. As shown in Fig.~\ref{fig:qpsk-results}(c), both $N$ and $\Delta f T_s$ strongly affect the BER performance of FLE. Increasing these parameters improves subprotocol detectability, but excessive values lead to degradation of the primary protocol, presenting a trade-off to the end user. Importantly, there exists an overlap region where both the primary protocol and the FLE subprotocol achieve high performance. This is evident from the FLE parameter sweep in Fig.~\ref{fig:qpsk-results}(d). If $\Delta f T_s$ is too small, the subprotocol signal is lost in the noise floor; if $N$ is too low, the PLL cannot reliably lock and track the induced frequency changes. Optimal performance is achieved through a balanced combination of these two parameters.

We present the following as a design rule. We first take the intersection of the primary protocol and subprotocol stability regions and then apply a conservative linear fit to the boundary. This places the operating envelope at
\[
0.0016 \le \Delta f T_s \le 0.0012N + 0.0080
\qquad \text{for } N \ge 12
\]

Using standard ITU channel models (EPA, EVA, EVB)\cite{itur_m1225_1997}, preliminary evaluations showed that more complex propagation environments—including frequency-selective fading and multipath effects—did not significantly degrade subprotocol performance. In several harsher channel conditions, the overlay even outperformed the underlying protocol. This behavior is attributed to the longer effective symbol duration, which improves robustness against multipath propagation, as well as the fact that the overlay shifts the entire base signal bandwidth uniformly, making localized frequency-selective impairments less pronounced. As a result, this work focuses primarily on the AWGN channel model.
 
\subsection{Over-the-Air Tests}

For the over-the-air experiments, operating parameters were selected based on the simulation results: a QPSK primary protocol with an FLE overlay using $N = 32$ and $\Delta f T_s = 0.0096$. Experiments were conducted at transmitter--receiver separations of 1~m and 20~m. These experiments serve as a preliminary feasibility study, not a full tactical-channel validation. At the shorter distance, near-perfect reception was observed for both the primary protocol and the FLE subprotocol. At the increased separation, the system continued to exhibit excellent performance, with both layers achieving measured BERs on the order of $10^{-6}$.

Although the observed BERs were ultimately determined by the SNR conditions of the two scenarios, the key result is that the primary protocol and the subprotocol exhibited similar error-rate trends and operational ranges. These experiments demonstrate that the secondary signaling layer can closely track the reliability of the underlying communication system while remaining effectively transparent to the primary protocol. In tested configurations, the subprotocol achieved data rates exceeding 900~kbps, with performance ultimately limited by the computational capability of the laptops running GNU Radio Companion rather than by the FLE mechanism itself.

\section{Conclusion}

This paper introduces FLE, a protocol-agnostic secondary signaling technique that can be used to embed spectrum coordination information into an existing wireless transmission through small, controlled frequency offsets. 

We analyzed FLE for a single-carrier system and characterized the trade-offs between secondary data rate and impact on the primary protocol. Simulation and over-the-air experiments demonstrate that FLE can be reliably decoded with negligible degradation to primary performance, achieving bit-error rates on the order of $10^{-6}$ for both links. These results show that excess synchronization margin can be repurposed as a lightweight, ad hoc signaling channel, enabling spectrum coordination between unassociated devices without requiring protocol interoperability, network association, or centralized infrastructure---properties well-aligned with the demands of tactical and coalition operations.


\bibliographystyle{IEEEtran}
\bibliography{references}

\end{document}